\documentclass[fleqn,10pt]{wlscirep}
\usepackage[utf8]{inputenc}
\usepackage[T1]{fontenc}
\usepackage{soul}
\title{Data-driven quantification and visualization of resilience metrics of power distribution system}

\author{Dingwei Wang}
\author{Salish Maharjan}
\author{Junyuan Zheng}
\author{Liming Liu}
\author[1,*]{Zhaoyu Wang}
\affil{Iowa State University, Department of Electrical and Computer Engineering, Ames, Iowa, 50010, USA}

\affil[*]{\textit{Corresponding Author}, email: wzy@iastate.edu}


\begin{abstract}

This paper presents a data-driven approach for quantifying the resilience of distribution power grids to extreme weather events using two key metrics: (a) the number of outages and (b) restoration time. The method leverages historical outage records maintained by power utilities and weather measurements collected by the National Oceanic and Atmospheric Administration (NOAA) to evaluate resilience across a utility's service territory. The proposed framework consists of three stages. First, outage events are systematically extracted from the outage records by temporally and spatially aggregating coincident component outages. In the second stage, weather zones across the service territory are delineated using a Voronoi polygon approach, based on the locations of NOAA weather sensors. Finally, data-driven models for outage fragility and restoration time are developed for each weather zone. These models enable the quantification and visualization of resilience metrics under varying intensities of extreme weather events. The proposed method is demonstrated using real-world data from a US distribution utility, located in Indianapolis, focused on wind- and precipitation-related events. The dataset spans two decades and includes over 160,000 outage records. 
\end{abstract}
\begin{document}

\flushbottom
\maketitle
\thispagestyle{empty}

% \noindent Please note: Abbreviations should be introduced at the first mention in the main text – no abbreviations lists. Suggested structure of main text (not enforced) is provided below.

\section*{Introduction}

% The Introduction section, of referenced text\cite{Figueredo:2009dg} expands on the background of the work (some overlap with the Abstract is acceptable). The introduction should not include subheadings.

\begin{figure}[ht]
\centering
\includegraphics[width=\linewidth]{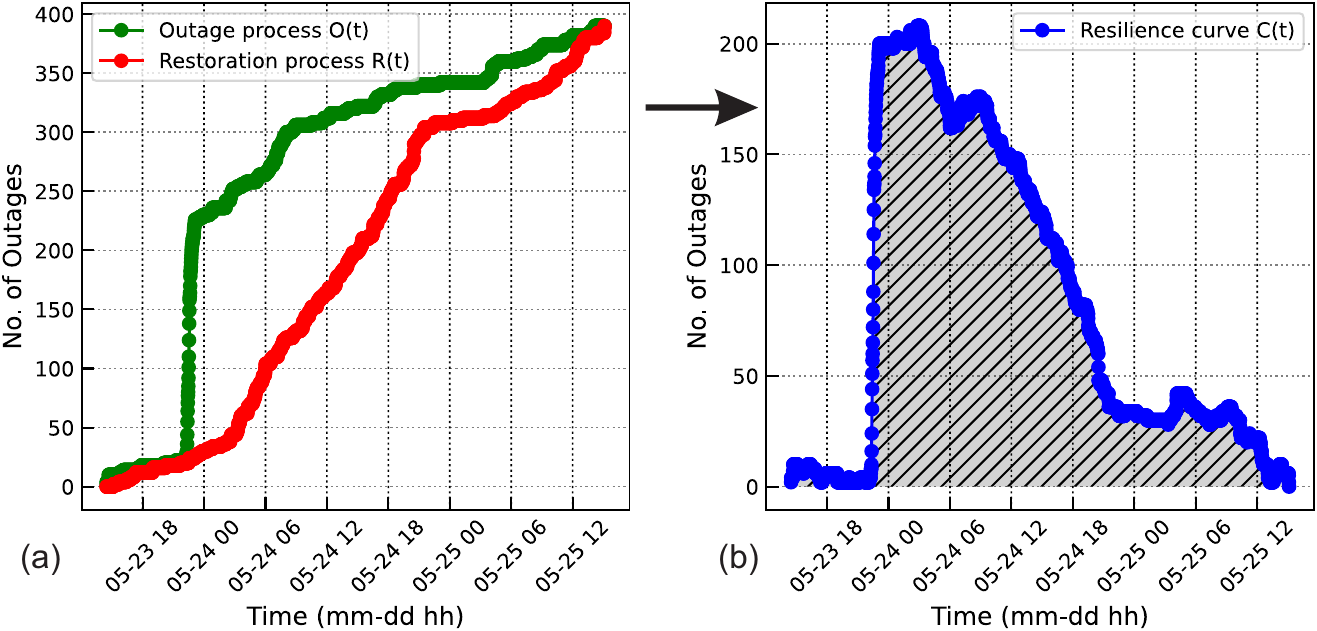}
\caption{(a) Outage accumulation ($O(t)$) and restoration process ($R(t)$) visualization during extreme wind event in Indianapolis. (b) Visualization of grid resilience ($C(t)$)}
\label{fig:outages}
\end{figure}

The power distribution systems are responsible for delivering electricity from high-voltage transmission networks to end-use customers through an interconnected network of substations, transformers, overhead and underground lines. Component outages occur randomly within the system and are often caused by factors such as aging infrastructure, animal interference, vegetation contact with power lines, or weather conditions. Among them, severe weather events are a significant contributing factor that can compound the number of outages in a short period of time \cite{res1}, leading to massive economic disruption. The impact of severe weather events on the power system can cause millions of dollars in economic loss \cite{extremeweathercost,outagelost2}. Once extreme weather subsides, the restoration process begins with the dispatch of repair crews to identify, isolate, and repair outages \cite{res2}. The speed and efficiency of this restoration depend on multiple factors, such as weather conditions, the number and size of the outages, the available crews and equipment, access difficulties, and situational awareness of the utility \cite{outagelost1}. However, the restoration rate always lags behind the outage rate. The outage accumulation process and the restoration process during extreme weather events can be studied leveraging the time-stamped records of component outages and restoration through the utility's outage management system (OMS). A real-case example of outage accumulation and utility response during an extreme event is illustrated in {Figure}~\ref{fig:outages}(a). The green curve represents the rapid accumulation of outages ($O(t)$) over a short period of time, which is characteristic of the impacts of extreme events. In contrast, the red curve shows the corresponding restoration progress ($R(t)$), which typically lags behind the outage rate. By the end of the event, all outages were fully restored. The resilience curve ($C(t)$) is obtained by subtracting the number of outages from restored ones (i.e., $C(t)=O(t)-R(t)$), which is shown in {Figure}~\ref{fig:outages}(b). The shaded area under the resilience curve is inversely related to grid resilience: the smaller the area, the higher the resilience. However, calculating this area accurately is a challenge. Therefore, we propose two practical resilience metrics: (a) the number of outages and (b) the total restoration time. Minimizing these values corresponds to reducing the shaded area, indicating better resilience.
% Resilience in the context of distribution systems is defined as the ability of the power grid to prepare for, absorb, adapt to, and rapidly recover from adverse events, including natural disasters and extreme weather disturbances \cite{res4,res5,kulkarni2024enhancing}. In Figure~\ref{fig:outages}, a steeper restoration curve that closely follows the outage curve indicates higher resilience, reflecting prompt and effective recovery, while a wider gap between the curves suggests a delayed response and increased vulnerability. Among various outage causes, such as equipment failure and animal contact, severe weather events, including high winds, flooding, and ice storms, remain the most dominant and unpredictable contributors to widespread outages. During such events, the restoration process is further complicated by the cumulative number of simultaneous outages and the real-time coordination of limited repair resources. Understanding these complex interdependencies is necessary for utilities seeking to improve restoration efficiency and enhance overall system resilience in the face of rapid climate change.

In recent years, several works have been made in quantifying the resilience of distribution systems to extreme weather events. These approaches can be broadly grouped into two categories based on their primary focus: \textit{Group I – Quantification of Fragility and Resilience Metrics} and \textit{Group II – Outage Restoration Modeling}. \textit{Group I – Quantification of Fragility and Resilience Metrics:} Fragility models characterize how system components or aggregated areas respond to increasing weather stress. Xu~\cite{xu2008undergrounding} developed exponential and power-law fragility models for poles and conductor spans under simulated hurricane conditions, estimating failures as a function of wind speed. Dunn~\cite{dunn2018fragility} and Reed~\cite{reed2008extremewinds} constructed empirical and analytical fragility curves from real-world outage data during high-wind events, identifying threshold wind intensities beyond which failure rates rise rapidly. Donaldson~\cite{donaldson2023resilience} and Murray~\cite{murray2014wind} correlated regional wind gusts with observed fault patterns in the UK, using normalized fragility metrics to compare geographic performance. In particular, Bjarnadottir~\cite{bjarnadottir2013hurricane} incorporated hurricane intensity and pole design into log-normal fragility modeling for long-term resilience assessment under climate change. In parallel, studies have extended fragility modeling to precipitation-driven events, particularly floods and rainstorms. Liu~\cite{liu2008spatialGLM} employed spatially generalized linear mixed models to relate hurricane and ice storm outages to precipitation, protection device density, and land features. Al Mamun~\cite{almamun2023resilience} and Davidson~\cite{davidson2003carolina} investigated the combined effects of rainfall, land cover, and storm surge on outage patterns, demonstrating that precipitation intensity significantly correlates with failure rates in flood-prone areas. Carrington~\cite{carrington2021metrics} and Ahmad~\cite{ahmad2024utilitydata} introduced resilience event extraction methods and area-level outage rate curves that quantify the aggregate system response to localized weather measurements, including wind and precipitation. \textit{Group II – Outage Restoration Modeling:} Restoration modeling under adverse weather conditions has focused on optimizing the timing, sequencing, and deployment of recovery resources. Arif~\cite{arif2018restoration} formulated a two-stage stochastic optimization problem to co-optimize network operation and crew routing under uncertain repair times, which is particularly relevant during storm-induced flooding, where access and damage assessment are delayed. Tan~\cite{tan2018hardening, tan2019scheduling} proposed mixed-integer linear programs to coordinate hardening and repair actions that minimize unserved energy, incorporating weather-triggered failures and restoration priorities. Jaech~\cite{jaech2019realtime} developed machine learning models to predict restoration durations using features such as cause codes, customer impact, and weather classifications. Cerrai~\cite{cerrai2019storm} and Kezunovic~\cite{kezunovic2018integration} utilized decision trees and logistic regression to integrate real-time precipitation and vegetation data to predict the probability of storm outages at fine geographic resolutions. These approaches demonstrate research on rainfall-related disruptions, which often involve cascading effects such as water intrusion, landslides, and saturated terrain conditions that complicate field response. In contrast to these resource-intensive simulations, Ahmad~\cite{ahmad2024utilitydata} presented a data-driven alternative by modifying historical outage records to reflect faster or earlier restoration scenarios, allowing utilities to rapidly estimate resilience improvements from operational improvements.

Despite the valuable progress achieved in existing studies, several important limitations remain that impact their effectiveness and practical relevance. First, many models analyze each outage record as an individual component-level incident, focusing on separate failures and restorations without accounting for the cumulative effect of multiple outages occurring within a short period in the same area. In reality, extreme weather events often result in groups of outages that occur closely in both time and space, placing significant pressure on utility resources. When the number of affected components increases rapidly in a concentrated region, restoration becomes more difficult due to the limited availability of repair crews and equipment. By overlooking the combined nature of these events, existing models may underestimate the full operational impact and provide incomplete assessments of system resilience. Second, most studies do not incorporate detailed spatial analysis of outage data. Service territories are often modeled as uniform areas, with no consideration for localized variations in exposure, terrain, or sensor coverage. Even when spatial zones are used, they are commonly defined as regular grids with fixed dimensions, which do not align with the actual distribution of weather monitoring stations or the geographical features of the infrastructure. This lack of spatial precision limits the ability of the models to reflect localized differences in outage behavior and risk. Third, although some studies attempt to divide systems into resilience zones, their zoning strategies are often based on artificial partitions that do not correspond to the spatial patterns of wind or precipitation. Moreover, most approaches apply a single model framework to both wind-related and precipitation-related outages, despite the fact that these two types of events arise from different causes. Wind events are typically associated with physical damage from falling trees or pressure on poles and wires, while precipitation events are more often linked to water accumulation, soil saturation, and equipment exposure. Using a common model for both types of events may reduce the accuracy of predictions and overlook important event-specific patterns. Finally, a large part of the literature relies on machine learning methods that, although capable of capturing complex relationships, require extensive training data, detailed preprocessing steps, and careful adjustment of model parameters. These requirements can create barriers to adoption in utility environments, where interpretability, computational efficiency, and ease of use are critical. Models that are difficult to understand or maintain may not be suitable for real-time decision-making or for use by operators without specialized technical expertise. These limitations highlight the need for a practical, data-driven framework that can capture the aggregated nature of outages, reflect spatial and weather-related variability, and support straightforward, reliable use in operational settings.

The outage restoration time modeling in existing resilience studies also presents several critical limitations that affect both its transparency and its applicability in operational contexts. First, many existing models estimate restoration performance using simple summary metrics such as average outage duration or mean time to restore. These models typically do not express the restoration time as a function of specific weather intensities or differentiated weather zones. As a result, they cannot predict how restoration dynamics vary under different levels of storm severity or in distinct geographic areas. In particular, restoration time is rarely expressed as a function of specific event intensity levels or spatially defined weather zones. This issue is especially pronounced in machine learning-based approaches, where outage records serve as inputs, and restoration durations are produced as outputs without explicit linkage to event characteristics. Although these models may achieve acceptable accuracy, they function as black boxes, providing little insight into how weather conditions or outage characteristics influence restoration performance. As a result, they offer limited interpretability and are difficult to validate or adapt to new operational conditions. Second, restoration curves in the literature are often constructed without accounting for the influence of outage volume on restoration dynamics. In severe weather events, the number of outages can increase dramatically in a short time, leading to saturation of crews and equipment, significantly slowing the recovery pace. Traditional models tend to assume a static or average restoration rate, failing to reflect the delays caused by high outage density, limited access, or constrained logistic coordination. This lack of sensitivity to outage clustering weakens the ability of models to simulate real restoration behavior during critical events. Third, restoration models commonly adopt the same structure for all weather-related outages, although field conditions and recovery constraints differ between event types. Treating these events with a single restoration formulation overlooks important event-specific patterns and limits the ability of models to support different outage scenarios. These limitations highlight the need for restoration models that are transparent, spatially contextualized, responsive to outage density, and tailored to the unique characteristics of different weather events.

Based on the identified limitations in the current literature, this work aims to provide a fully data-driven, comprehensive, interpretable, and spatially resolved framework for distribution system resilience quantification and outage restoration modeling under extreme weather events. The main objectives, motivations, and contributions of the paper are as follows: First, a framework for delineating the utility's service territory using the Voronoi polygon method based on National Oceanic and Atmospheric Administration (NOAA) sensor locations is developed to form weather zones. Subsequently, overlaying geospatial mapping of outage records enabled spatially differentiated resilience analysis across these zones. Second, an event extraction method is developed to group temporally and spatially coincident outages, enabling the modeling of cumulative impacts during high-impact weather events. This approach overcomes the limitation of treating outages as isolated incidents, providing a more accurate representation of system resilience. Third, data-driven fragility models are constructed to relate the number of outages in each event to the corresponding weather intensity within defined zones, offering interpretable and zone-specific failure predictions under varying hazard conditions. Finally, restoration time models are formulated by correlating the total restoration duration with the number of outages in each event, capturing the effect of outage volume on restoration performance. Note that separate modeling frameworks are developed for wind and precipitation events, addressing the inaccuracy introduced by applying a single unified model to distinct weather types with fundamentally different operational characteristics.

\section*{Methods}
\begin{figure}[htbp]
\centering
\includegraphics[width=1\textwidth]{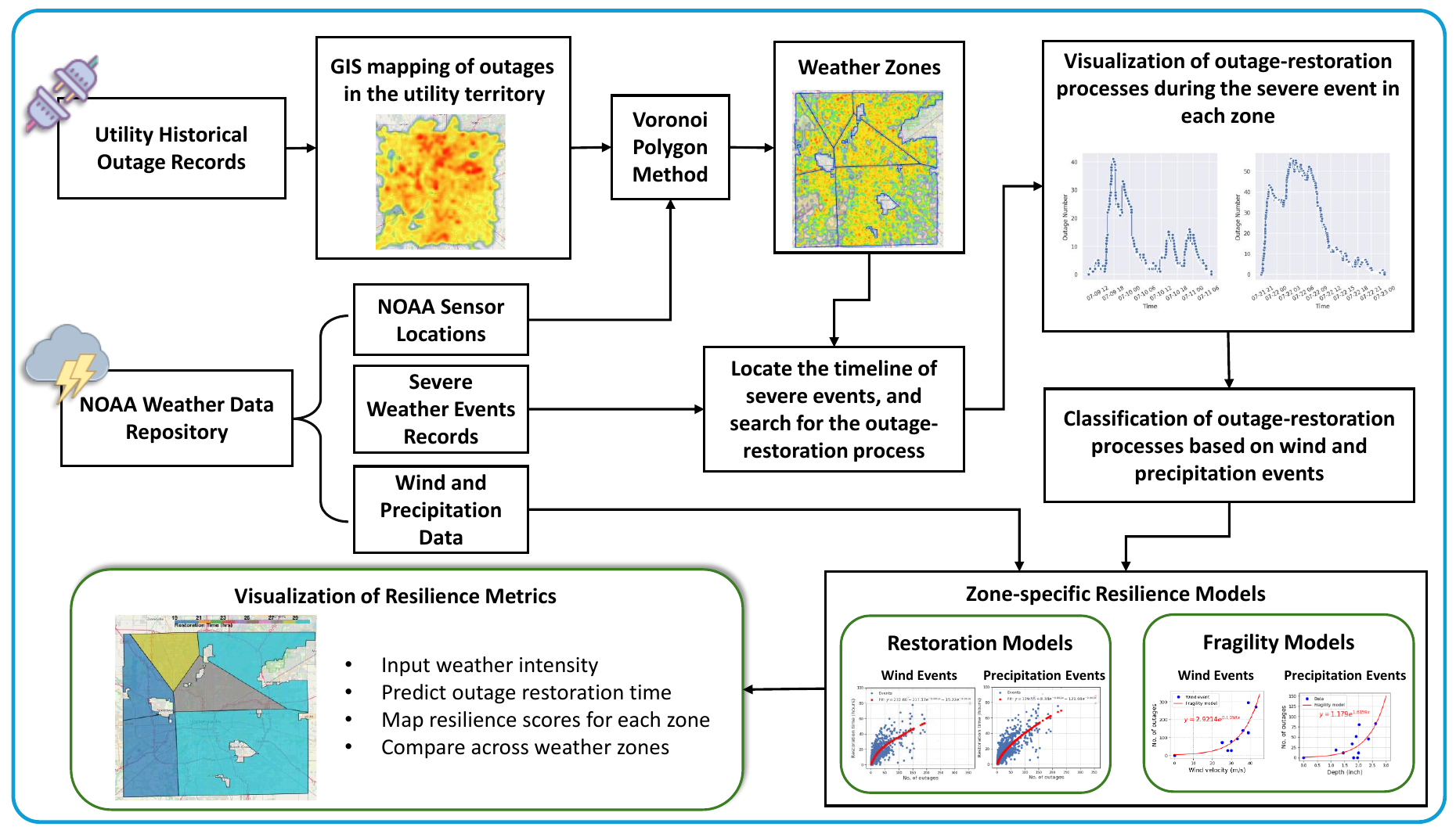}
\caption{Flowchart of the proposed data-driven resilience quantification and visualization framework.}
\label{fig: flow}
\end{figure}

This study proposes a data-driven framework for quantifying the resilience of power distribution systems to extreme weather by integrating utility outage data with meteorological observations. The overall workflow, shown in Figure~\ref{fig: flow}, transforms the raw operational and weather records into zone-specific fragility models, restoration time models, and spatial visualizations of the power distribution system resilience.

The process begins with data acquisition from two main sources: (i) historical outage records from the utility’s outage management system, containing timestamps, locations, cause codes, and customer interrupted, and (ii) NOAA weather data, including hourly sensor measurements and severe weather event records. Both datasets are preprocessed to remove missing or erroneous values. Following data preprocessing, GIS mapping of outage locations is performed to visualize the spatial distribution of historical device failures across the utility service territory. The Voronoi polygon method is applied to divide the service area into wind and precipitation zones, with each zone assigned to its nearest NOAA weather sensor.

The outage data are further organized into outage–restoration events, defined as a continuous period beginning with the first outage that appears and ending when all affected components are restored. This event-based structure captures the cumulative effect of simultaneous outages, which is essential to understand operational stress during major interruptions. The timeline of severe weather events serves as a guide for locating such events within the large-scale outage dataset. Severe weather event logs are then used to extract two subsets of these events: (i) wind events and (ii) precipitation events, based on temporal and location overlap with outage records. With these event datasets prepared, fragility models are developed to quantify the relationship between weather intensity and outage occurrence for each zone. Wind events are linked to wind speed, and precipitation events to precipitation depth, producing exponential functions that represent the expected number of outages at given weather conditions. In parallel, restoration time models are built using all outage–restoration events to relate outage volume to total restoration time, capturing how cumulative outage numbers influence recovery rates.

Finally, fragility and restoration models are integrated for resilience metric visualization. For a given weather condition, such as a specified wind velocity or rainfall depth, the fragility model estimates the number of outages in each zone, which the restoration model then translates into predicted restoration durations. These results are mapped across all zones to provide an intuitive and spatially resolved comparison of resilience, highlighting areas with relatively stronger or weaker recovery performance under specific weather conditions.

\section*{Results}
\subsection*{Outage and Weather Data Description}
This study utilizes two datasets: (i) outage records provided by a US distribution utility in Indianapolis, and (ii) open weather data and severe weather events from NOAA, including wind and precipitation measurements from multiple weather stations and all severe weather-related extreme event history within the utility’s service area \cite{noaa, noaasevere}. Both datasets span a 20-year period from 2004 to 2024 and are temporally aligned to support joint analysis.

The outage dataset contains a total of 168,462 individual outage records. Each record corresponds to a discrete outage that affects a component in the distribution system and includes the start and end times of the outage, accurate to seconds; the duration of restoration in minutes; the number of customers interrupted; and the associated cause code. Before proceeding with analysis, the dataset was carefully preprocessed to improve reliability and internal consistency. Due to equipment malfunctions or manual reporting errors, historical outage data often contains incomplete or inaccurate entries, which can lead to misleading conclusions in data-driven analysis. To address this, records with missing values in key fields, such as timestamps, restoration duration, or customer count, were discarded. In addition, entries with logically inconsistent values were filtered out. These included cases where the restoration time exceeded the overall outage duration or where the values fell outside reasonable operational limits. 
% This data cleaning process helps ensure that the final dataset reflects realistic system behavior and supports the development of robust and interpretable resilience models. 
% Given a big picture of the outage restoration, the restoration times in this dataset vary widely, with the majority of events restored within 20 to 200 minutes. Specifically, 52,047 outages were restored in the 50–100 minute range, while 31,799 outages required more than 200 minutes. The causes of the outage were classified to identify the key factors of system failure. The top three causes are overhead equipment failures, tree-related and animal-related, and weather-related events. 

The weather sensor data includes hourly wind and precipitation data from 12 stations across the study region. Wind-related features include hourly average wind speed and fastest 2-minute wind speed, all recorded in meters per second. Precipitation-related features include hourly totals of rainfall, snowfall, and ground-level snow depth, measured in inches. All sensor data were preprocessed to ensure temporal alignment with outage records. The severe weather event dataset \cite{noaasevere} is obtained from a separate NOAA database and consists of documented weather incidents classified by event type. Each record includes a start and end time, accurate to the minute, event type, starting location, and affected area of the event represented as longitude and latitude, and a brief description of the event. The types of events considered in this study include tornadoes, high-speed wind events (classified as wind events), snowstorms, and flooding (classified as precipitation events). However, unlike the weather sensor dataset, severe event records do not contain direct numerical measurements of wind speed or precipitation.

% After collecting and preprocessing datasets, each outage record is aligned with the corresponding weather sensor data based on its timestamp. Specifically, the start time of each outage is matched to the closest available weather observation from the NOAA dataset. Both wind and precipitation data are recorded on an hourly basis from NOAA stations. For each outage, the weather observation corresponding to the same hour as the outage start time is selected. \textcolor{red}{I am thinking put this alignment work to later subsections where we perform Geospatial Mapping of Outages. Because the OMS does not have weather records, that is why we need to align the weather data with mapped outages.}

% \textcolor{red}{Following data collection and preprocessing, each outage record was temporally aligned with the nearest available weather sensor data based on its start time. Specifically, the start time of each outage was used as a reference point to retrieve the corresponding hourly weather observations, such as wind speed and precipitation, from the closest available NOAA sensor station. In parallel, severe weather event records were used to identify and extract outage records that occurred under documented extreme weather conditions. For each severe event, outage records were extracted based on both spatial and temporal overlap. An outage was selected if its start time occurred within the start and end times of the documented event, and its location fell within the geographic boundaries associated with the reported incident.}

\subsection*{GIS Mapping of Component Outages}

\begin{figure}[htbp]
\centering
\includegraphics[width=0.6\textwidth]{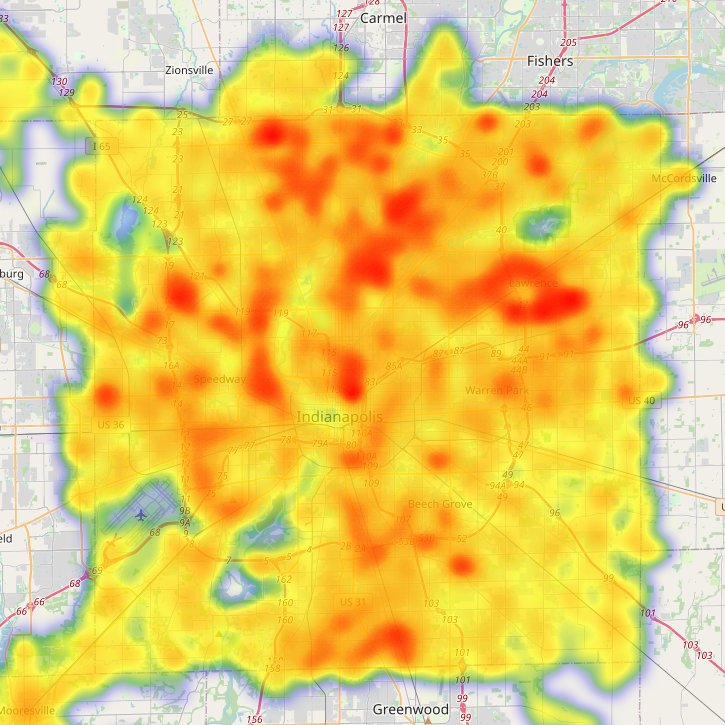}
\caption{Mapping of component outages during 2004 to 2024 on a GIS map.}
\label{fig: outage_mapping}
\end{figure}

To visualize the spatial distribution of the outages in the utility service area, each outage record from the OMS was mapped to its corresponding geographical location. The OMS database provides information on damaged components or protective devices, such as tripped breakers or blown fuses that triggered each outage. Using the unique identifiers for these components, their GIS coordinates were retrieved and plotted to generate a heatmap of the outage density. Figure~\ref{fig: outage_mapping} presents the resulting heatmap, where the regions with higher outage concentrations are shown in red and yellow. This geospatial representation demonstrates clusters of vulnerability within the service territory. However, it is important to note that OMS records do not contain any weather measurements or meteorological context, the weather dataset mentioned in the previous subsection needs to be aligned to each OMS record.

\subsection*{Outage-restoration Events Extraction}
To systematically analyze outage accumulation and restoration behavior, this work defines \textit{outage-restoration events} based on the temporal progression of component outage records. Rather than independently evaluating each outage, outages are grouped into events that represent continuous disturbance intervals followed by full system recovery. As discussed in the Introduction section, restoration performance is not solely determined by the characteristics of a single outage, but is heavily influenced by the number and timing of coincident outages within a given time frame. When multiple outages occur in rapid succession, often triggered by severe weather, the restoration process becomes increasingly complex due to limited repair resources, unavoidable delays, and difficulty in arrangement. By aggregating temporally overlapping outages into outage events, this process will analyze how restoration time scales with the number of concurrent outages and provide a more realistic representation of system behavior under high system stress conditions.

\begin{figure}[htbp]
\centering
\includegraphics[width=\textwidth]{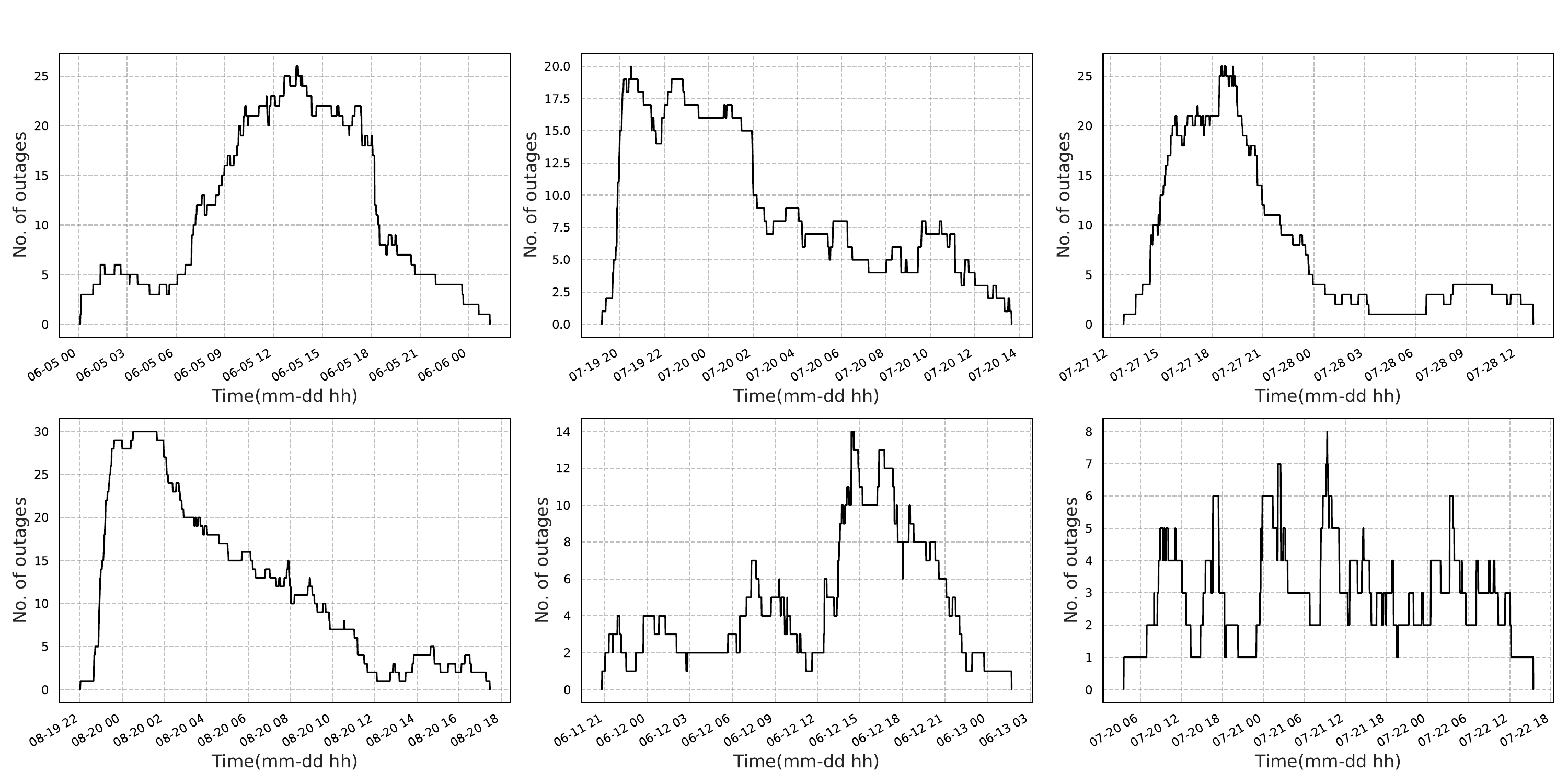}
\caption{Instances of extracted outage-restoration events during extreme weather.}
\label{fig: event}
\end{figure}

An event begins with the first outage that occurs when all components are functional and ends when all outages have been restored, bringing the cumulative number of active outages, denoted as $C(t)$, back to zero. Specifically, the process involves sorting all outage start and restoration times in chronological order and tracking $C(t)$ at each time point. The value of $C(t)$ increases when a new outage appears and decreases when an outage is restored. Each time $C(t)$ returns to zero, the current event concludes, and any subsequent outage marks the start of a new event. For example, if an event includes $n$ outage onsets, it must also contain $n$ corresponding restorations to ensure that $C(t)$ returns to zero at the end of the event window. This approach captures the cumulative behavior of coinciding outages and allows each event to be characterized by two primary features: the number of outages and the total restoration time, defined as the event restoration time from the first outage to the final restoration within the event. Figure \ref{fig: event} illustrates representative examples of extracted outage-restoration events. These plots illustrate how outages accumulate and are subsequently resolved during each event. The curves typically exhibit a sharp increase in the outage count at the beginning of an event, followed by a gradual decline as restorations are completed. The variation in shapes and durations reflects the diversity of event profiles, including both short, concentrated disruptions and longer, more complex restoration processes.

\subsection*{Delineation of Weather Zone via Voronoi Polygon Method}

To capture the spatial variation in weather conditions across the service territory, the studied area is delineated into localized weather zones based on the distribution of NOAA weather stations. Specifically, two wind and six precipitation stations located within the region are used to distinguish wind and precipitation weather zones for resilience analyses. Each weather station serves as a reference point for one zone, ensuring that the weather data associated with each outage are drawn from the most geographically relevant source.

The division is accomplished using the Voronoi polygon method, a spatial partitioning technique that assigns every point in the region to its nearest weather station. Given a set of weather stations ${s_1, s_2, \dots, s_n}$ with known geographic coordinates, the Voronoi cell for station $s_i$ consists of all points in the plane that are closer to $s_i$ than to any other station $s_j$, where $j \neq i$. This results in a tessellation of the area into non-overlapping polygons, each representing the area of influence of a particular weather station. The method is implemented using the QGIS software platform, which supports geospatial processing and visualization of shape files based on station coordinates. The resulting wind and precipitation zones defined for this study area are presented in the Results section.

\begin{figure}[htbp]
\centering
\includegraphics[width=\linewidth]{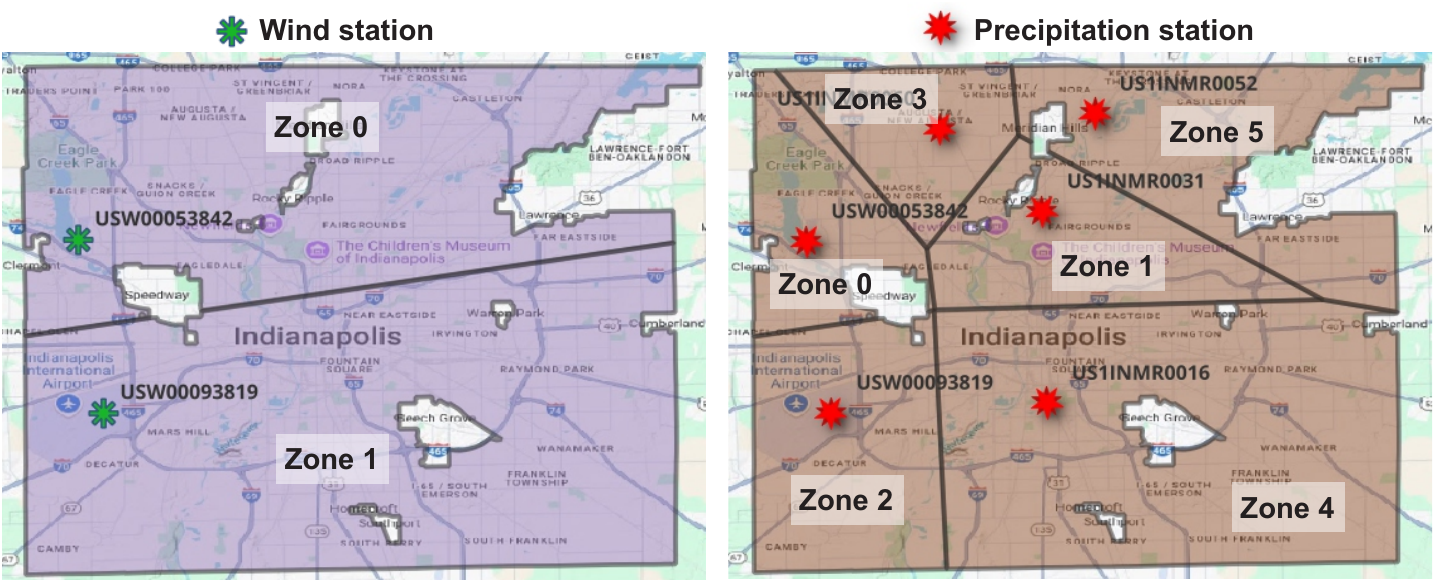}
\caption{Division of the city of Indianapolis into weather zones using the Voronoi polygon method. (Left) Two wind zones based on wind station locations. (Right) Six precipitation zones based on precipitation station locations. Each polygon defines the area closest to a given weather station.}
\label{fig: zones}
\end{figure}

Based on the spatial distribution of NOAA weather sensors and the measurement types available at each station, two wind zones and six precipitation zones were defined for the utility service region using the Voronoi polygon method. Wind zones are determined using the locations of stations that record wind-related features, such as hourly wind averages and gust speeds. The precipitation zones are defined using the locations of sensors that provide hourly rainfall, snowfall, and snow depth. The defined wind and precipitation zones are shown in Figure~\ref{fig: zones}. The left part shows the division of the service territory into two wind zones, each surrounding one of the two wind stations. The right part shows the division into six precipitation zones, each linked to one of the six precipitation stations. 

\subsection*{Data-driven Resilience Models}
This section introduces two models to quantify resilience across the distribution utility's service territory. The first is a fragility model that correlates the number of outages with weather variables. The second is a restoration time model that relates the restoration time to the number of outages. These resilience metrics are subsequently quantified using real-world data from a utility serving the city of Indianapolis. 

\subsubsection*{Data-driven Fragility Models Using Severe Weather Outage Events}

A structured, multi-step process was followed to analyze system resilience against severe weather events. Each step was carefully designed to ensure the integration of weather records, outage data, and spatially defined weather zones into meaningful fragility models.

The first step involves compiling a dataset of historical severe weather records affecting the service area between 2004 and 2024. This dataset includes records classified by location, duration, and weather types such as high winds, tornadoes, floods, heavy snow, and thunderstorms. In the second step, all severe weather records are classified into two categories: wind-related events (e.g., tornadoes, high winds) and precipitation-related events (e.g., floods, snowstorms). This step classifies and associates the relevant weather variables, such as wind speed or precipitation levels, with each type of event. After that, each weather event is assigned to a defined weather zone based on the spatial coverage of the weather stations and the Voronoi polygon boundaries established earlier. Each event is matched to the zone in which it occurred, which supports the development of fragility curves reflecting the unique conditions of each zone. Then, the number of outages during each weather event is computed. This is done by filtering the outage dataset by event time window and geographic location, capturing only the outages that occurred within the event duration and the corresponding weather zone. In addition, we link each event to its corresponding weather measurements. For wind-related events, wind speed observations using the fastest 2-minute averages are extracted. For precipitation-related events, the hourly precipitation, snowfall, and snow depth are obtained. These values are aligned with the event temporal window to represent the intensity of the event. Finally, the collected data are aggregated at the zone level, yielding paired observations of weather intensity and outage impacts. These data are then used to construct fragility models, which express the number of outages as a function of weather severity. Separate fragility models are developed for wind and precipitation zones using wind and precipitation events, as their purpose is to characterize the relationship between weather intensity (e.g., wind speed or precipitation depth) and the number of outages resulting from those weather conditions. Exponential functions are used for model fitting, as they capture the nonlinear increase in outage likelihood with rising weather intensity.

% The outcomes included two distinct sets of fragility curves: For the defined wind weather zones, curves were developed to show (1) the number of outages vs. wind speed. For the precipitation weather zones, curves were created for (2) the number of outages vs. precipitation levels. 
% An example of the developed fragility model is shown in Figure~\ref{fig: fragility example}. This figure illustrates the relationship between wind velocity and the number of outages recorded during wind-related events. The blue dots represent the observed outage counts associated with individual wind events, and the red curve shows an exponential fitting function of the form $y = 0.0002e^{0.3675x}$, where $x$ is wind velocity (m/s) and $y$ is the number of outages. The model captures the increasing outage likelihood as wind speed increases. \textcolor{red}{Still thinking how to deal with this part and the example fig.}

% \begin{figure}[htbp]
% \centering
% \includegraphics[width=0.75\linewidth]{fragility example.png}
%     \caption{An example fragility model fitted to wind event data.\textcolor{red}{Seems like the example plot is redundant, might need to remove. This contains duplicate information in the results section.}}
% \label{fig: fragility example}
% \end{figure}

\subsubsection*{Data-driven Restoration Time Models Using Outage Events}

To characterize restoration performance across varying weather conditions, restoration time models are developed using the full set of outage-restoration events. These models aim to capture the general relationship between the number of outages and the total restoration time, regardless of the underlying cause, and therefore reflect the operational response characteristics across all types of outage conditions.

For each extracted event, as discussed in Section "Outage-restoration Events Extraction", two quantities are summarized: the total number of outages and the total restoration time. These values are plotted in scatter diagrams to explore the empirical relationship between the number of outages and the duration of restoration. Following this, events are grouped by weather zone, and separate restoration models are fitted for each zone. A combination of constant and double exponential functions is selected for its flexibility in capturing both the initial increase and the saturation effects observed in large-scale events.

% Figure~\ref{fig: restoration model example} shows an example of a restoration time model fitted to outage event data. The blue dots represent observed restoration durations for events with different outage sizes. The red curve represents the fitted model, which represents the relationship of restoration time given the number of outages in an event.
% \textcolor{red}{Same to Fig. 8. Do we need this? Or is there any way to draw a different figure to correspond the text.}

% \begin{figure}[htbp]
% \centering
% \includegraphics[width=0.75\linewidth]{restoration example.png}
%     \caption{An example of the Restoration Time Model.}
% \label{fig: restoration model example}
% \end{figure}

This method produces a model for estimating restoration times across wind and precipitation zones using the number of outages predicted by the fragility models. By linking outage volume to restoration time in an event-specific manner, this method supports data-driven assessment of restoration performance under different severe weather conditions.

\subsubsection*{Fragility and Restoration Time Models for Wind Events}
\begin{figure}[htbp]
\centering
\includegraphics[width=\linewidth]{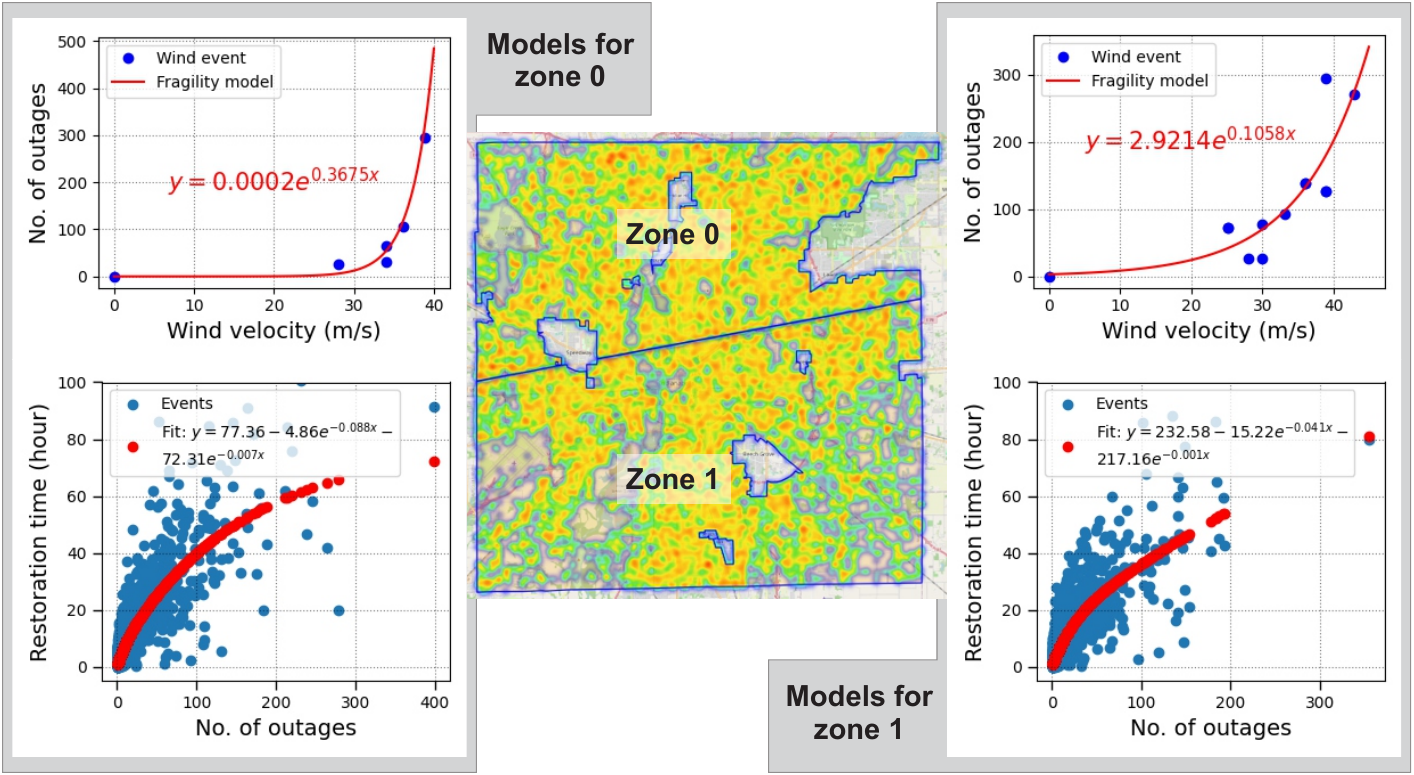}
\caption{Fragility and restoration time models for Wind Zones 0 and 1. The central map shows the spatial boundary of the two wind zones in the studied area.}
\label{fig: wind_models}
\end{figure}
To account for the impacts of wind events on the distribution system, zone-specific fragility and restoration time models are developed. Figure~\ref{fig: wind_models} displays the results for Wind Zone 0 and Wind Zone 1. The central map shows the spatial division of the two wind zones, with background shading representing the outage density throughout the study area. Wind Zone 0 covers the northern part of the region, while Wind Zone 1 covers the southern part.

The fragility model for Wind Zone 0 is shown in the upper left. The blue dots represent observed outages for individual wind events, plotted against maximum wind velocity (in m/s). The red curve shows the best-fit exponential function of the form $y = 0.0002e^{0.3675x}$, where $x$ is the wind speed and $y$ is the number of outages. The curve demonstrates a rapid increase in outage counts once wind speeds exceed approximately 28 m/s, which is a threshold effect where system vulnerability intensifies significantly. This pattern reflects the relationship between outage numbers and wind speed in this zone. The corresponding restoration model for Wind Zone 0 is shown in the bottom left. Similarly, each blue dot represents a historical outage event, with the x-axis indicating the number of outages and the y-axis indicating the total restoration time in hours. The red curve represents the fitted model, a two-term exponential function that captures both the initial rapid increase in restoration time and the eventual saturation effect. This relationship shows that, as the outage volume increases, restoration becomes delayed, likely due to constraints in available repair crews and resource allocation.

The upper right panel presents the fragility model for Wind Zone 1. Compared to Zone 0, this zone exhibits a slightly different fragility profile. The fitted exponential model can be expressed as $y = 2.9214e^{0.1058x}$, and this curve shows a more gradual increase in the outage count with wind speed. This indicates that Zone 1 may have relatively lower resilience against wind-related failures, or possibly differences in terrain or vegetation contributing to this trend. The figure on the bottom right shows the restoration model for Wind Zone 1. As in Zone 0, the restoration time increases nonlinearly with the outage volume, but the fitted curve here suggests a sharper increase at lower outage counts. This pattern may indicate less operational redundancy or more constrained access conditions in Zone 1. The model is defined by the expression $y = 232.60 - 217.17e^{-0.001x} - 15.22e^{-0.041x}$.

\subsubsection*{Fragility and Restoration Time Models for Precipitation Events}
\begin{figure}[htbp]
\centering
\includegraphics[width=0.9\textwidth]{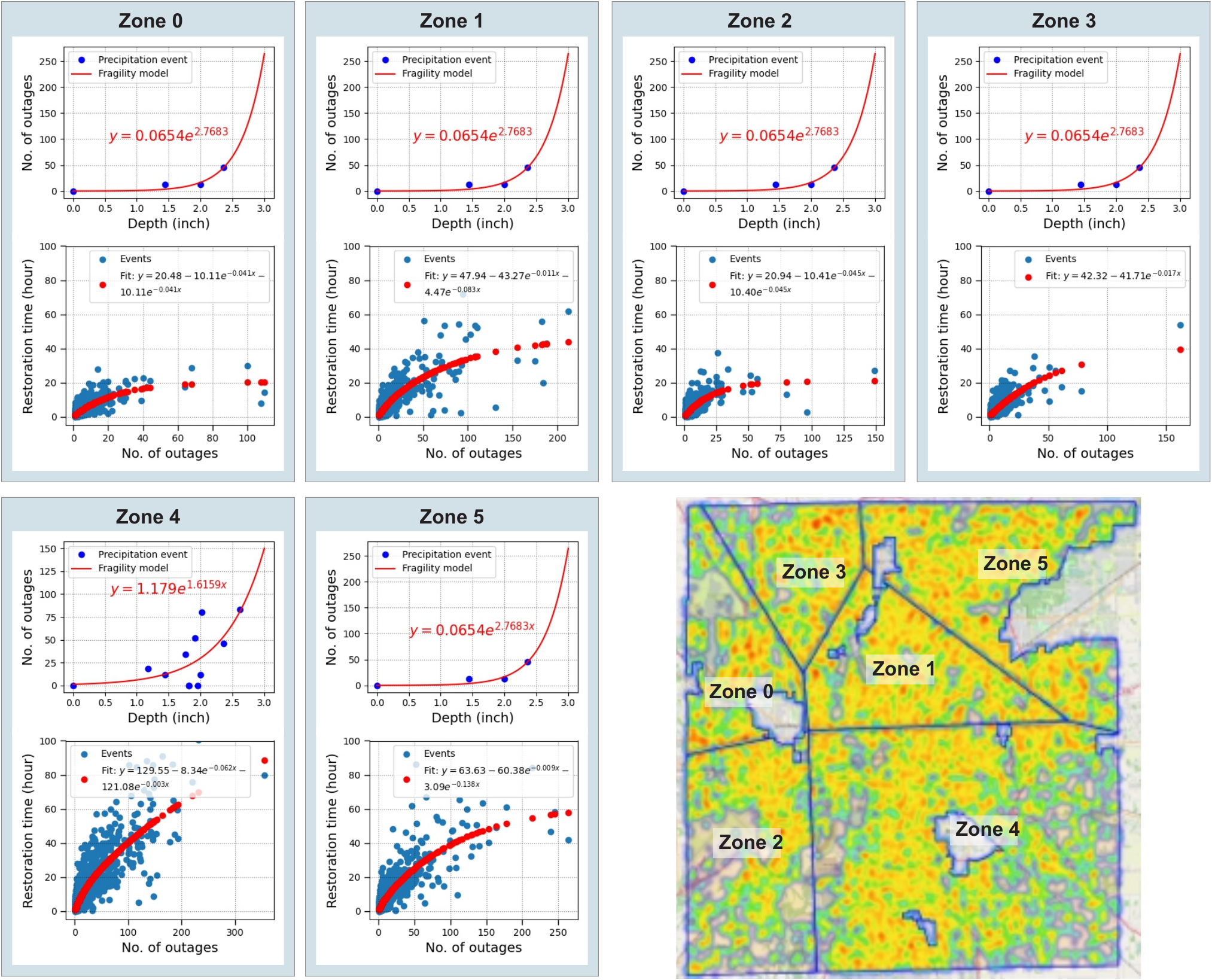}
\caption{Fragility and restoration time models for Precipitation Zones 0 through 5. The map illustrates the spatial boundaries of the precipitation zones.}
\label{fig: precip_models}
\end{figure}

Indianapolis has six precipitation zones, and Figure~\ref{fig: precip_models} presents the fragility and restoration models for each zone. The map in the lower right illustrates the boundaries of the six precipitation zones, and the background is the density of the outage records. In the fragility plots, blue dots represent observed outage counts during individual precipitation events, plotted against precipitation depth (in inches). The red curves correspond to the fitted exponential fragility models. The corresponding restoration time models are shown below or next to the fragility plots. Each blue point represents an outage event, with the $x$-axis showing the number of outages and the $y$-axis indicating the total restoration time in hours. The red curves represent fitted double-exponential functions. Although exact coefficients vary by zone, the general trend in all cases shows that restoration time increases nonlinearly with outage volume.

A notable observation is that Zones 0, 1, 2, 3, and 5 all share the same fragility function: $y = 0.0654e^{2.7683x}$. These results do not imply identical physical characteristics across these zones, but instead reflect the structure of the precipitation event data. These five zones are geographically close to each other and are frequently affected by the same precipitation events. Because zones are defined based on proximity to NOAA precipitation stations using the Voronoi method, a single large-scale precipitation event, such as a widespread rainstorm, can impact multiple zones simultaneously. As a result, the precipitation records and associated outage patterns used for fragility modeling in these zones are similar. This overlap in data leads to the convergence of fitted fragility curves across Zones 0 through 3 and Zone 5. In other geographic areas with more localized or spatially varied rainfall patterns, where precipitation events do not affect multiple zones simultaneously, it is expected that the fragility curves would diverge and reflect the unique characteristics of each zone.

Zone 4 shows a distinct fragility curve, modeled as $y = 1.179e^{1.6159x}$, which grows more slowly than the fragility functions in the other zones. This may indicate lower sensitivity to precipitation depth, possibly due to reduced exposure, fewer vulnerable components, or differences in system configuration. The restoration model for Zone 4 also shows a steeper increase in duration with increasing outage count, showing longer recovery periods under stress conditions. For Zone 5, while the fragility model is identical to that of Zones 0–3, the restoration curve deviates. The fitted model indicates a slower increase in restoration time at lower outage volumes, followed by a steeper climb once the number of outages exceeds approximately 100. This behavior may reflect the threshold of repair resources.

% The case study conducted in this paper uses real-world outage and weather data from the utility service territory in Indiana to evaluate the proposed resilience quantification framework. The analysis covers a 20-year period of outage records from 2004 to 2024. The outage dataset contains over 168,000 individual records, while the weather dataset includes hourly wind and precipitation observations collected from 8 NOAA weather stations within the region. Based on the spatial distribution of these stations, the study area is divided into wind zones and precipitation zones using the Voronoi polygon method. For each identified outage-restoration event, the number of outages, restoration duration, and associated weather conditions are extracted and aligned. The fragility curves and restoration time models developed in this study are applied to this dataset to demonstrate and quantify the resilience of the distribution system. The following subsections present analyzed outage records, weather zone definitions, and fitted fragility and restoration models, along with the resulting resilience metrics under various wind and precipitation conditions.

\subsection*{Visualization of Resilience metrics}
\begin{figure}[htbp]
\centering
\includegraphics[width=\linewidth]{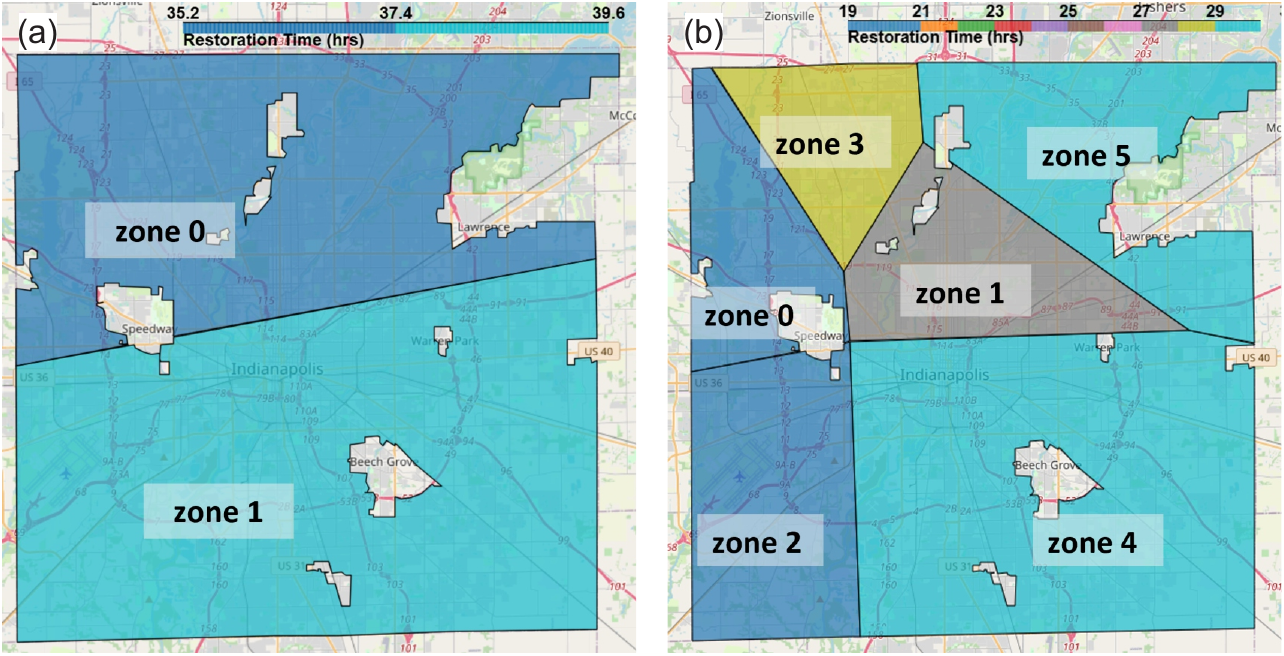}
\caption{(a) Visualization of restoration time across Wind Zones 0 and 1 for a wind event with velocity of 35 m/s. (b) Visualization of predicted restoration time across precipitation zones for a 2.5-inch precipitation event.}
\label{fig:visualize_resilience}
\end{figure}

The developed fragility and restoration time models can be combined to generate zone-specific predictions of outage restoration time under a given weather scenario. For each target weather condition, such as wind velocity (m/s) for a wind event or precipitation depth (inches) for a precipitation event, the fragility model is used first to estimate the expected number of outages in each zone. Specifically, the wind or precipitation level is input into each zone's corresponding exponential fragility model, yielding an estimated outage count for that scenario. The predicted number of outages is then input into the corresponding restoration time model. The restoration model maps the outage count to an expected total restoration time. This two-step process estimates the restoration time in each zone, given a particular level of wind or precipitation. Applying this procedure across all zones generates a complete set of restoration time estimates for the entire service area. The results are visualized on a spatial map, with each zone shaded according to its predicted restoration time under the specified weather conditions. These visualizations directly compare resilience between zones and help identify areas more vulnerable to specific weather conditions.

Figure~\ref{fig:visualize_resilience}(a) presents an example of a wind event with a velocity of 35~m/s. In this visualization, darker shading indicates shorter predicted restoration durations and thus higher resilience. As shown in the figure, Wind Zone 0 exhibits a shorter expected restoration time compared to Zone 1 under the same wind intensity, indicating a stronger resilience performance. Figure~\ref{fig:visualize_resilience}(b) shows the restoration performance for a precipitation event with 2.5 inches of rainfall. As the map illustrates, Zones 0 and 2 demonstrate relatively shorter restoration times, indicating higher resilience to precipitation-driven outages, while Zones 4 and 5 exhibit longer restoration durations and greater vulnerability. This spatial comparison can help utilities target specific zones for resilience improvement efforts.

\section*{Discussion}

The framework presented in this study offers a utility-oriented approach to resilience quantification that emphasizes accessibility, clarity, and ease of implementation. Unlike many recent methods that rely on complex simulation environments or machine learning models requiring large-scale data training and parameter tuning, the proposed approach builds directly on outages and weather records that utilities already maintain. This makes the tool inherently practical and adaptable, requiring minimal technical overhead or advanced modeling expertise.

One of the primary advantages of this method is its interpretability. Both the fragility and restoration time models are constructed using straightforward relationships that clearly show how weather intensity and outage volume relate to system impacts. These models are intentionally designed to be transparent and editable, enabling utility engineers and planners to understand, validate, and explain the model outputs without relying on opaque or black-box techniques. This aligns with a core requirement from utilities: the ability to use data-driven tools that support planning and operational decision-making without introducing additional uncertainty or abstraction.

Another key benefit is that the framework is tailored to real-world operational conditions and constraints. The method accounts for spatial variation by using geographically defined zones based on available weather station coverage, allowing utilities to compare resilience across service areas using localized metrics. The output is not only quantitative but also actionable, offering insights that can support targeted hardening, outage planning, and resource allocation. Importantly, all components of the tool are built around data types already collected in routine operations, eliminating the need for additional instrumentation or external modeling dependencies.

Despite its practical strengths, the method does have limitations. Its accuracy depends on the quality and completeness of historical data, and the models may be less reliable in areas or scenarios where historical events are sparse. Furthermore, while the zoning approach captures spatial variation in weather exposure, it does not incorporate differences in infrastructure conditions, vegetation, or network configuration that may affect outage probability and restoration effort. Future enhancements could involve integrating asset-specific data or incorporating logistical variables such as crew travel time and repair constraints to improve restoration model precision.

In general, the proposed method provides a practical and scalable foundation for resilience assessment in distribution systems. This work contributes a scalable and sustainable tool for advancing resilience assessment in modern distribution systems by establishing the method in operational data and aligning it with utility planning needs.

\bibliography{sample}

% \noindent LaTeX formats citations and references automatically using the bibliography records in your .bib file, which you can edit via the project menu. Use the cite command for an inline citation, e.g.  \cite{Hao:gidmaps:2014}.

% For data citations of datasets uploaded to e.g. \emph{figshare}, please use the \verb|howpublished| option in the bib entry to specify the platform and the link, as in the \verb|Hao:gidmaps:2014| example in the sample bibliography file.

\section*{Acknowledgements}

The work was funded and supported by the U.S. Department of Energy Joint Office of Energy and Transportation under DE-EE0011234.

\section*{Author contributions statement}

S.M. and Z.W. conceived the experiments,  D.W., L.L., and J.Z. conducted the experiments, D.W. and S.M. analyzed the results. D.W. and S.M. prepared the manuscript, and all authors reviewed it.

\section*{Data Availability}
The data that support the findings of this study are available from the corresponding author upon reasonable request.

\section*{Code Availability}

The code packages that support the findings of this study are available from the corresponding author upon reasonable request.

\section*{Competing interests}

The authors declare no competing interests.

\end{document}